\documentclass[prd,twocolumn,showpacs,showkeys,preprintnumbers,amsmath,amssymb]{revtex4-1}
\usepackage{graphicx}
\usepackage{dcolumn}
\usepackage{wrapfig} 
\usepackage{bm}
\usepackage{float}
\begin{document}

\preprint{CERN-PH-TH/2013--063, LYCEN 2013-02}\vspace*{0.5cm}

\title{Anomaly mediated supersymmetric models and Higgs data from the LHC}
\author{Alexandre Arbey$^{1,2,3}$}
\email{alexandre.arbey@ens-lyon.fr}
\author{Aldo Deandrea$^{2}$}
\email{deandrea@ipnl.in2p3.fr}
\author{Farvah Mahmoudi$^{3,4}$}
\email{mahmoudi@in2p3.fr}
\author{Ahmad Tarhini$^{2}$}
\email{tarhini@ipnl.in2p3.fr}
\affiliation{\vspace*{0.2cm}$^{1}$CRAL, Observatoire de Lyon, 
9 avenue Charles Andr\'e, F-69561 Saint-Genis Laval Cedex, France;\\ 
CNRS, UMR5574; Ecole Normale Sup\'erieure de Lyon, France.\\
$^{2}$Universit\'e de Lyon, France; Universit\'e Lyon 1, CNRS/IN2P3, UMR5822 IPNL, F-69622~Villeurbanne Cedex, France.\\
$^3$ CERN Theory Division, Physics Department, CH-1211 Geneva 23, Switzerland.\\
$^4$ Clermont Universit\'e, Universit\'e Blaise Pascal, CNRS/IN2P3, LPC, BP 10448, F-63000 Clermont-Ferrand, France
 }
 
\begin{abstract}
Anomaly mediation models are well motivated supersymmetry breaking scenarios which appear as alternatives to the minimal 
supergravity paradigm. These models are quite compelling from the theoretical point of view 
and it is therefore important to test if they are also viable models for phenomenology.
We perform a study of these models in the light of all standard flavour, collider and dark matter 
constraints, including also the recent Higgs boson measurements for the mass and signal strengths in the different decay 
channels. The minimal Anomaly Mediated Supersymmetry Breaking (AMSB) scenario can satisfy in part of its parameter space the dark matter requirement but is only marginally consistent with the current Higgs boson mass value. 
The HyperCharge-AMSB and Mixed Moduli-AMSB scenarios can 
better describe present data from dark matter, flavour, low energy physics and are consistent with the measured mass of the Higgs 
boson. The inclusion of the preferred signal strengths for the Higgs boson decay channels shows that for $\tan \beta \gtrsim 5$ the HyperCharge-AMSB and Mixed Moduli-AMSB models can be consistent with the present Higgs boson data. 
In contrast the minimal AMSB has a narrower allowed range in $\tan \beta$. These different AMSB scenarios, while consistent 
with present Higgs boson measurements, can be further tested by future more precise data in the Higgs sector. 
\end{abstract}
\maketitle

\section{Introduction}

The ATLAS and CMS experiments at the Large Hadron Collider (LHC) have reported the discovery of a new boson compatible with the 
Standard Model (SM) Higgs in July 2012 \cite{ATLAS:2012zz,CMS:2012zz}, and updated results of the measurements of the Higgs 
couplings with more precision have been recently released 
\cite{ATLAS-2012-160,ATLAS-2012-161,CMS-12-043,CMS-12-044,ATLAS-2012-168,ATLAS-2012-169} and 
\cite{ATLAS-CONF-2013-014,CMS-PAS-HIG-2013-002,ATLAS-CONF-2013-012,CMS-PAS-HIG-2013-001, 
ATLAS-CONF-2013-013,CMS-PAS-HIG-2013-002,ATLAS-CONF-2013-030,CMS-PAS-HIG-2013-003}. All the results are compatible 
with the predictions for a SM Higgs boson with a mass of about 126 GeV. This discovery is especially important in the context of 
new physics models, and in particular supersymmetry (SUSY), where the Higgs mass and decay rates can be related to the SUSY 
parameters. 

In this paper, we consider specific scenarios in which anomaly mediation supersymmetry breaking mechanisms are assumed. In 
particular, we discuss the implications of $B$ physics data, LHC Higgs measurements and cold dark matter relic abundance. We 
discuss different possibilities, such as minimal anomaly mediation (mAMSB) \cite{Randall:1998uk,amsb}, hypercharged anomaly mediation (HC-AMSB) \cite{Dermisek:2007qi} and
mixed moduli-anomaly mediation (MM-AMSB) \cite{Choi:2005uz}. Anomaly 
mediation models in relation with dark matter and cosmology were also discussed in \cite{Arbey:2011gu}.

Anomaly Mediation Supersymmetry Breaking (AMSB) models are based on the fact that the conformal 
anomaly gives a 
general and model independent contribution to gaugino masses which is always present and which can be the dominant contribution 
when there is no direct tree level coupling which transfers the SUSY breaking from the hidden sector to the observable sector.
This contrasts with the mSUGRA mechanism which is based on the existence of specific tree-level terms. These models are 
theoretically very appealing as based on existing forces (gravity) and on well motivated string theory ideas. A well-known problem however in the 
AMSB scenario is the presence of tachyonic sleptons. This problem can be solved 
assuming the presence of an intermediate threshold scale. Different supersymmetry breaking scenarios can be obtained starting from 
this framework, such as mAMSB, HC-AMSB and MM-AMSB, which we briefly review in the following. It is quite remarkable that present 
data from different sectors and in particular the recent LHC results for the Higgs boson are able to partially constrain the parameter 
space of these models. We discuss in the following the implications of such data and the interplay with other constraints coming 
principally from $B$-physics and dark matter relic abundance.

This paper is organised as follows. In section \ref{sec:models} we discuss briefly the theoretical framework of the different 
anomaly mediated supersymmetry breaking set-ups. In section \ref{sec:constraints} the implications of the flavour physics and relic 
density constraints are presented, as well as the Higgs 
mass constraints and the possibilities to obtain branching ratios for the light CP-even Higgs in agreement with the present results at the 
LHC for the parameter spaces of the different AMSB models. 
Conclusions are given in section \ref{sec:conclusion}.

\section{Theoretical framework}
\label{sec:models}
\subsection{mAMSB}
\label{sec:model_mAMSB}
The minimal AMSB (mAMSB) scenario can be described both from a higher-dimensional spacetime point 
of view \cite{Randall:1998uk}, where the SUSY breaking occurs on a separate brane and is communicated to the visible sector via super-Weyl anomaly, or from a four dimensional perspective \cite{amsb}, where a model-independent contribution to the
gaugino mass is obtained from the conformal anomaly. In models without singlets this mechanism is the dominant one in the gaugino mass. The soft SUSY breaking terms can be calculated in terms of a single 
parameter, the gravitino mass $m_{3/2}$. Nonetheless, as an attempt to avoid the tachyonic slepton problem, it is assumed that the 
scalar particles have a universal mass $m_0$ at the GUT scale, which leads to positive AMSB soft SUSY
breaking terms. The mAMSB model possesses only four parameters:
\begin{equation}
 m_0, m_{3/2}, \tan \beta, \mbox{sign}(\mu) \;.
\end{equation}
The soft SUSY breaking gaugino mass terms are related to $m_{3/2}$ by \cite{Baer:2010kd}:
\begin{equation}
 M_i = \frac{\beta_i}{g_i} m_{3/2}\;,
\end{equation}
where $i=1\cdots3$, $g_i$ are the coupling constants and $\beta_i$ the corresponding $\beta$ functions. The soft SUSY breaking sfermion mass terms and fermion trilinear couplings are given by:
\begin{eqnarray}
 m_{\tilde{f}}^2 &=& -\frac{1}{4}\left(\sum_{i=1}^3 \frac{d\gamma}{dg_i}\beta_i +\frac{d\gamma}{dy_f}\beta_f \right) m_{3/2}^2 + m_0^2 \;,
  \label{msfmamsb}
 \\
 A_f &=& \frac{\beta_f}{y_f} m_{3/2}\;,
\end{eqnarray}
where $\beta_f$ is the $\beta$ function corresponding to the Yukawa coupling $y_f$ and $\gamma = \partial \ln Z /\partial \ln \mu$, where 
$Z$ is the wave function renormalisation constant.

\subsection{HC-AMSB}
\label{sec:model_HC-AMSB}
A substitute approach to solve the tachyonic lepton problem is the hypercharge anomaly mediated supersymmetry breaking 
(HC-AMSB) scenario \cite{Dermisek:2007qi}. In this model, the MSSM is bound to a D-brane 
and a geometrically separated hidden sector generates a hypercharge gaugino mass \cite{Baer:2009wz}. 
Therefore, the tachyon problem can be solved by an increase in the slepton masses which results from an additional
contribution to the gaugino mass $M_1$. We parametrise the HC-AMSB symmetry breaking using a dimensionless quantity $\alpha$ 
which determines the hypercharge contribution relative to the soft bino mass term in AMSB.

The HC-AMSB scenario has also four parameters:
\begin{equation}
 \alpha = \frac{\tilde{M}_1}{m_{3/2}}, m_{3/2}, \tan \beta, \mbox{sgn}(\mu) \;,
\end{equation}
where $\tilde{M}_1$ is the additional hypercharge contribution to $M_1$. 

The anomaly mediation and hypercharge mediation have a common theoretical set-up, and the two types of mediation are able to 
compensate the phenomenological shortcomings of each other, giving rise to a realistic and well motivated model. 
Indeed the minimal AMSB model predicts a negative mass squared for the sleptons (and 
prefers heavy squarks) while the pure hypercharge mediation suffers from negative squared masses for stops and 
sbottoms (and prefers heavy sleptons). Combining the hypercharge and anomaly mediation set-ups gives rise to a 
phenomenologically viable spectra in a sizeable range of the parameter space of the model. 

In the HC-AMSB model, the soft SUSY breaking terms are identical to the AMSB ones, apart from the bino and fermion mass terms \cite{Baer:2010kd}:
\begin{eqnarray}
 M_1 &=& \left( \alpha + \frac{\beta_1}{g_1} \right) m_{3/2}\;,\\
 m_{\tilde{f}}^2 &=& -\frac{1}{4}\left(\sum_{i=1}^3 \frac{d\gamma}{dg_i}\beta_i +\frac{d\gamma}{dy_f}\beta_f \right) m_{3/2}^2 \;.
 \label{msfhcamsb}
\end{eqnarray}

At the two loop level, the other gaugino masses, $M_{2}$ and $M_{3}$, receive a contribution from the bino mass term~\cite{Baer:2009wz}.

\subsection{MM-AMSB}
\label{sec:model_MM-AMSB}
The Mixed Modulus Anomaly mediated SUSY breaking (MM-AMSB) scenario \cite{Choi:2005uz} can be used as a third possibility to 
solve the tachyon problem. It is based on type-IIB superstrings with stabilised moduli \cite{Baer:2006id}. In this model, the moduli fields 
which describe the extra dimensions and the Weyl anomaly have comparable contributions to the SUSY breaking in the 
observable sector.  The spatial extra dimensions are compactified with flux which brings 
to a minimum in the potential of moduli and represents a starting point to find the fundamental state which leads to MSSM at low 
energy \cite{Baer:2006id}.
The soft SUSY breaking terms receive contributions of comparable magnitude from both anomaly and modulus, which can increase the 
slepton masses and solve the tachyon problem. The MM-AMSB scenario has four parameters:
\begin{equation}
 \alpha , m_{3/2}, \tan \beta, \mbox{sgn}(\mu) \;,
\end{equation}
where $\alpha$ parametrises the relative contributions of modulus mediation and anomaly mediation to the soft breaking terms. 
A large $\alpha$ corresponds to a mediation from the moduli, and a small $\alpha$ to a mediation from the anomaly.
Indeed in the limit where $\alpha \rightarrow$ 0, we obtain pure AMSB soft SUSY breaking terms with a negative squared mass for the 
sleptons. For intermediate values of $\alpha$ which are more interesting for our studies, the problem of tachyonic sleptons is absent 
\cite{Choi:2005uz}. 
The mass scale of supersymmetry breaking parameters is given by the gravitino mass $m_{3/2}$. The soft SUSY parameters are given by \cite{Baer:2006id}
\begin{eqnarray}
 M_i &=& \left(\frac{\alpha}{16\pi^2} + \frac{\beta_i}{g_i} \right) m_{3/2}\;,\\
 m_{\tilde{f}}^2 &=& \left\{ \frac{\alpha^2}{256 \pi^4} + \frac{\alpha}{4 \pi^2} \xi_f  -\frac{1}{4}\left(\sum_{i=1}^3 \frac{d\gamma}{dg_i}\beta_i +\frac{d\gamma}{dy_f}\beta_f \right) \right\}m_{3/2}^2 \;,\nonumber\\
 A_f &=& \left(- \frac{3 \alpha}{16\pi^2} + \frac{\beta_f}{y_f}\right) m_{3/2}\;,\nonumber
\end{eqnarray}
with
\begin{equation}
\xi_f=\frac{3}{4}y_f^{2}- \sum_a g_a^2 C_2^a (f)\;,
\end{equation}
where $C_2^a$ and $g_a$ are the quadratic Casimir and coupling of the $a^{\rm th}$ gauge group corresponding to the sfermion.


\section{Tools and Constraints}
\label{sec:constraints}

In order to study the different AMSB scenarios, we used {\tt ISAJET 7.82} \cite{isajet} to generate the SUSY spectra, compute the flavour 
observables and relic density with {\tt SuperIso Relic v3.2} \cite{superiso,superiso_relic}, and we calculate the Higgs branching fractions 
and decay widths with {\tt HDECAY 5.11} \cite{hdecay}. In the following, we disregard the case of negative $\mbox{sign}(\mu)$ since it 
is disfavoured by the muon anomalous magnetic moment constraint, when assuming that supersymmetric contributions fill the gap 
between the measurements and the SM predictions (typically squarks are assumed heavy to explain the Higgs boson mass and LHC 
bounds, while sleptons, neutralinos and charginos may be light to be consistent with the $g-2$ muon results, see \cite{Endo:2013bba} 
for a recent analysis). Also, we impose the condition on the SUSY breaking scale 
$M_S = \sqrt{m_{\tilde{t}_1} m_{\tilde{t}_2}} < 3$ TeV as a typical scale to limit fine-tuning.

\subsection{Flavour bounds}
\label{sec:flavour}
It is well known that flavour physics observables provide important indirect constraints on the MSSM as they are sensitive to the SUSY parameters through virtual corrections. Similar considerations apply also in the 
case of the models under study. 

We first consider the inclusive branching ratio of $B \to X_s \gamma$. This decay has been thoroughly studied in the literature as its SM contributions only appear at loop level. The theoretical uncertainties as well as the 
experimental errors are also very well under control. The $B \to X_s \gamma$ branching ratio is particularly constraining in the large $\tan\beta$ region where it receives large corrections from the SUSY loops.
We use the following interval at 95\% C.L.:
\begin{equation}
 2.63 \times 10^{-4} < \mbox{BR}(B \to X_s \gamma) < 4.23 \times 10^{-4} \;.
\end{equation}
which is obtained using the latest experimental world average from Heavy Flavor Averaging Group (HFAG) of $(3.43 \pm 0.21 \pm 0.07) \times 10^{-4}$ \cite{Amhis:2012bh}, after taking into account the theoretical and experimental errors \cite{superiso,Mahmoudi:2007gd}.

Another important observable in constraining SUSY parameters is the branching ratio of $B_s \to \mu^+ \mu^-$, which is also a loop level observable and suffers from helicity suppression in the SM. In SUSY it can 
receive extremely large enhancements by several orders of magnitude at large $\tan\beta$. The first evidence for this decay has been reported by the LHCb collaboration very recently 
\cite{Aaij:2012ct}. We use the following 95\% C.L. interval which includes 10\% theoretical error~\cite{Mahmoudi:2012un}:
\begin{equation}
 0.99 \times 10^{-9} < \mbox{BR}(B_s \to \mu^+ \mu^-)_{\tt untag} < 6.47 \times 10^{-9} \;,
\end{equation}
where {\tt untag} denotes the untagged branching fraction, which can be derived from the CP-averaged branching fraction and directly 
compared to the experimental measurement \cite{DescotesGenon:2011pb,DeBruyn:2012wk,Arbey:2012ax}.

The purely leptonic decay of $B_u \to \tau \nu$ on the other hand is sensitive to supersymmetry through the exchange of a charged 
Higgs boson already at tree level, which does not suffer from the helicity suppression of the SM contribution with the exchange of a 
$W$ boson. This decay can therefore provide stringent constraints.
The combination of the most recent Belle and Babar results gives $(1.14 \pm 0.23) \times 10^{-4}$ \cite{Adachi:2012mm,Lees:2012ju} which, including the theoretical errors, leads to the following allowed interval:
\begin{equation}
 0.40 \times 10^{-4} < \mbox{BR}(B_u \to \tau \nu) < 1.88 \times 10^{-4} \;.
\end{equation}
We used $f_B = 194\pm 10$ MeV \cite{Mahmoudi:2012un} and $V_{ub} =(4.15\pm 0.49)\times 10^{-3}$ \cite{Beringer:1900zz} for the calculation of this branching ratio.

Other flavour observables could be added to this list, however we have just included the most stringent ones for this analysis. A more 
complete analysis including all flavour information requires in principle a global fit to all observables, similar to the ones performed to 
test the Standard Model. This however goes beyond the scope of this preliminary screening of the extensions of AMSB models 
discussed here.

\subsection{Relic density}
\label{sec:relic}

The WMAP data \cite{Hinshaw:2012fq} provide precise observations of the cold dark matter density in the Universe. We use them to impose constraints on the AMSB parameter spaces by computing the relic density with {\tt SuperIso Relic}. We consider the WMAP 
interval at 95\% C.L. increased by 10\% of theoretical error \cite{Hindmarsh:2005ix,Baro:2007em} to account for the uncertainties in the calculation of the relic density:
\begin{equation}
0.068 < \Omega_\chi h^2 < 0.155 \;. \label{Oh2_tight}
\end{equation}

However the relic density constraint can be falsified in alternative cosmological model \cite{Kamionkowski:1990ni} or if dark matter is 
composed by more than one species (with {\it e.g} moduli \cite{Acharya:2008bk}, axions or axinos \cite{Steffen:2008qp,Baer:2011hx})
 and we therefore also consider a loose interval:
\begin{equation}
10^{-4} < \Omega_\chi h^2 < 0.155 \;, \label{Oh2_loose}
\end{equation}
in which we relaxed the lower bound.

In addition to these bounds, we impose the neutralino to be the lightest supersymmetric particle (LSP) to avoid the cosmological problems related to charged or not-so-weakly-interacting relics. 

\subsection{Higgs searches}

The discovery of a Higgs-like particle at the LHC provides important information on the 
MSSM~\cite{Arbey:2011ab,Hall:2011aa,Baer:2011ab,Heinemeyer:2011aa,Christensen:2012ei,Brummer:2012ns,Ghosh:2012dh,Fowlie:2012im,CahillRowley:2012rv,Arbey:2012dq,Buckley:2012em,Cao:2012yn,Antusch:2012gv,AbdusSalam:2012sy,CahillRowley:2012kx,Altmannshofer:2012ks,Arbey:2012bp,Okada:2012nr,Ellis:2012nv,Ibanez:2013gf,Nath:2013okv,Han:2013gba,Boehm:2013qva}. In the following, we associate the newly discovered boson to the lightest 
CP-even Higgs~$h$.
The Higgs mass value close to 126 GeV brings constraints on the parameter space of supersymmetric models which enter the radiative 
corrections. 
The leading part of these corrections arises from the top/stop loops:
\begin{eqnarray}
(\Delta M_h^2)_{\tilde t} \approx  \frac{3  G_F}{\sqrt{2} \pi^2}\, m_t^4 \left [ -\ln \left(\frac{m_t^2}{M_S^2} \right)+ \frac{X_t^2}{M_S^2} \left ( 1 - \frac{X_t^2}{12 M_S^2} \right )  \right ] \;, \;\; \;\;
\end{eqnarray}
where $M_S = \sqrt{m_{\tilde t_1} m_{\tilde t_2}}$ and $X_t = A_t - \mu/\tan\beta$ is the stop mixing parameter. This correction is maximised for $|X_t| = \sqrt{6} \, M_S$, which is referred to as ``maximal mixing" scenario, and minimised for  $X_t =0$ in the ``minimal mixing" scenario. The ``typical mixing" scenario corresponds to the intermediate values of $|X_t| \approx M_S$.

In the figures, the constraint on the Higgs boson mass will be taken at the two sigma level, $121.5 < M_h < 129.9$ GeV. 
The extra information provided by the measurements of Higgs branching ratios, provides extra useful constraints.
\begin{table}[!t]
\begin{center}
\begin{tabular}{|c|c|c|}
\hline
 & Value & Experiment \\ \hline \hline
$M_h$     & 125.7$\pm$2.1 GeV & ATLAS\cite{ATLAS-CONF-2013-014}, CMS\cite{CMS-PAS-HIG-2013-002} \\ 
$\mu_{\gamma \gamma}$ & 1.20$\pm$0.30 & ATLAS\cite{ATLAS-CONF-2013-012}, CMS\cite{CMS-PAS-HIG-2013-001} \\
$\mu_{Z Z}$ & 1.10$\pm$0.22 & ATLAS\cite{ATLAS-CONF-2013-013}, CMS\cite{CMS-PAS-HIG-2013-002} \\
$\mu_{W W}$ & 0.77$\pm$0.21 & ATLAS\cite{ATLAS-CONF-2013-030}, CMS\cite{CMS-PAS-HIG-2013-003} \\                                 
\hline
$\mu_{b \bar b}$  & 1.12 $\pm$ 0.45 & ATLAS\cite{ATLAS-CONF-2012-161}, CMS\cite{CMS-PAS-HIG-2012-044}, 
CDF,D0\cite{Aaltonen:2012qt}\\ 
$\mu_{\tau \tau}$ & 1.01 $\pm$ 0.36 & ATLAS\cite{ATLAS-CONF-2012-161}, CMS\cite{CMS-PAS-HIG-2013-004}\\ 
\hline
\end{tabular}
\end{center}
\caption{Experimental average for the Higgs mass and rates~\cite{Arbey:2013jla}.}
\label{tab:input} 
\end{table}
The latest LHC measurements of the Higgs mass and decay rates are summarised in Table~\ref{tab:input}. We use in the following the 
by now standard notation of signal strengths normalised to the SM expectation, defined as:
\begin{eqnarray}
\mu_{\gamma\gamma,VV} &=& \frac{\sigma({\rm gluon\;fusion})}{\sigma_{\rm SM}({\rm gluon\;fusion})} \; \frac{{\rm BR}(h \to \gamma\gamma,VV)}{{\rm BR}_{\rm SM}(H \to \gamma\gamma,VV)}\;\;\;\;\\
\mu_{\tau\tau} &=& \frac{\sigma({\rm VBF})}{\sigma_{\rm SM}({\rm VBF})} \; \frac{{\rm BR}(h \to \tau\tau)}{{\rm BR}_{\rm SM}(H \to \tau\tau)}\;,\\
\mu_{b\bar{b}} &=& \frac{\sigma({\rm HV})}{\sigma_{\rm SM}({\rm HV})} \; \frac{{\rm BR}(h \to b\bar{b})}{{\rm BR}_{\rm SM}(H \to b\bar{b})}\;,
\end{eqnarray}
where $VV$ refers to vector boson $ZZ$ or $WW$ production, and VBF and HV stand for vector boson fusion and associated Higgs 
vector boson production. For the $\mu_{\gamma\gamma}$ signal strength note that we use the average from ATLAS and CMS as a 
guideline, but this should be taken with some care as the two experiments have quite different central values, ATLAS has 
$\mu_{\gamma\gamma}=1.65^{+0.34}_{-0.30}$ while CMS reports $0.78\pm 0.27$.
Hereafter, we do not show the constraints from $\mu_{ZZ}$, as they are very similar to the ones from the $WW$ channel.

To evaluate the Higgs production cross sections normalised to the SM values, we use:
\begin{equation}
\frac{\sigma({\rm gluon\;fusion})}{\sigma_{\rm SM}({\rm gluon\;fusion})} \approx \frac{\Gamma^h}{\Gamma_{\rm SM}^H} \; \frac{{\rm BR}(h \to gg)}{{\rm BR}_{\rm SM}(H \to gg)}\;,
\end{equation}
\begin{equation}
\frac{\sigma({\rm VBF})}{\sigma_{\rm SM}({\rm VBF})} \approx \frac{\sigma({\rm HV})}{\sigma_{\rm SM}({\rm HV})} \approx \frac{\Gamma^h}{\Gamma_{\rm SM}^H} \; \frac{{\rm BR}(h \to VV)}{{\rm BR}_{\rm SM}(H \to VV)}\;,
\end{equation}
where $\Gamma^h$ and $\Gamma_{\rm SM}^H$ are respectively the MSSM $h$ and SM $H$ total decay widths.

In the following, we do not impose strict intervals on the calculated signal strengths, but we comment on the compatibility of the results 
with the experimental data.

\section{Results}
\begin{figure*}[ht]
\begin{center}
\begin{tabular}{cc}
\includegraphics[width=0.422\textwidth]{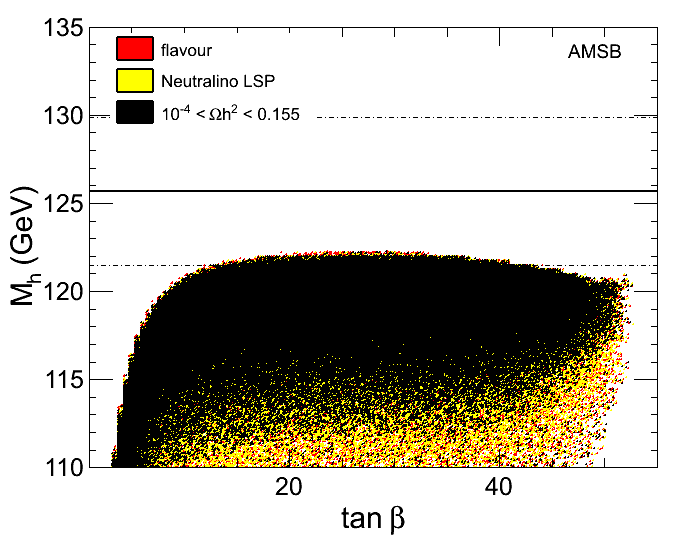}&
\includegraphics[width=0.422\textwidth]{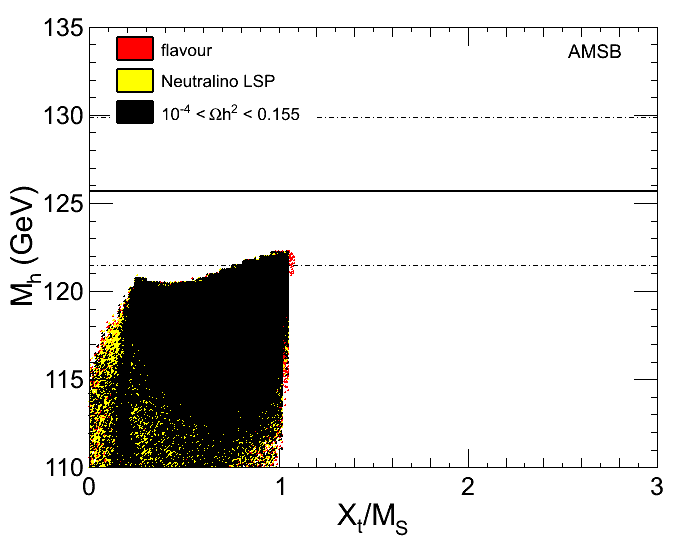}
\end{tabular}
\end{center}
\caption{Light Higgs mass as functions of $\tan\beta$ (left panel) and $X_t/M_S$ (right panel) in mAMSB. The red points are all points compatible with the constraints from flavour physics described in Sec.~\ref{sec:flavour}. The yellow points have also a neutralino LSP. The black points are in addition consistent with the upper bound of the relic density constraint. The horizontal solid line corresponds to the central value of the Higgs mass and the dashed lines to the 2$\sigma$ deviations.\label{mh0_amsb}}
\end{figure*}

\begin{figure*}[bht]
\begin{center}
\begin{tabular}{cc}
\includegraphics[width=0.422\textwidth]{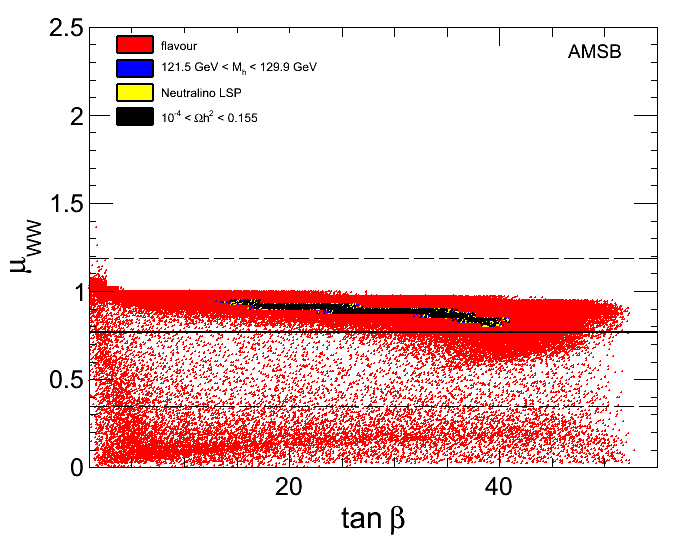}&
\includegraphics[width=0.422\textwidth]{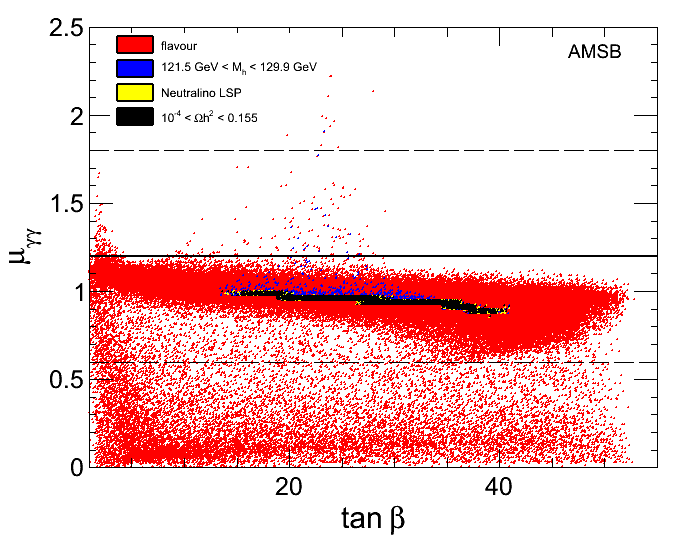}\\
\includegraphics[width=0.422\textwidth]{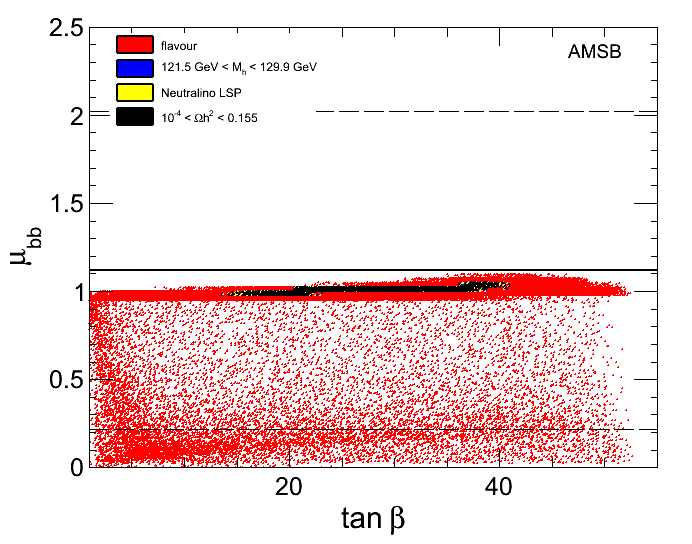}&
\includegraphics[width=0.422\textwidth]{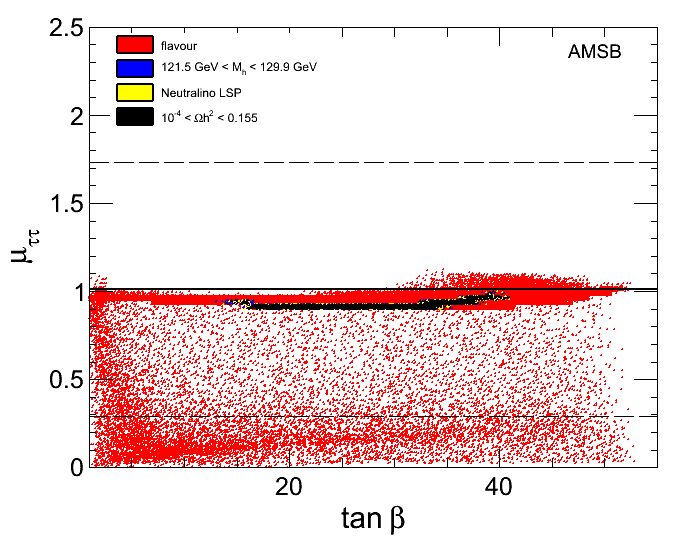}
\end{tabular}
\end{center}
\caption{\label{BR_tanb_amsb}$\mu_{WW}$ (upper left), $\mu_{\gamma\gamma}$ (upper right), $\mu_{b\bar{b}}$ (lower left) and $\mu_{\tau\tau}$ (lower right) as functions of $\tan\beta$ in the mAMSB model. The red points are favoured by the flavour physics constraints, the blue points are compatible with the Higgs mass constraint, the yellow points have a neutralino LSP and the black points in addition are compatible with the upper bound of the relic density constraint. The yellow and blue regions almost coincide with the black one, so most yellow and blue point are masked by the black region. The horizontal solid lines correspond to the experimental central values given in Table~\ref{tab:input} and the dashed lines to the $2\sigma$ intervals.}
\end{figure*}
We consider the constraints from flavour physics, dark matter and LHC Higgs searches in the context of minimal AMSB, hypercharge AMSB and mixed-moduli AMSB. We show in the following how the available parameter space is reduced in these different models 
when applying the available constraints.

\subsection{mAMSB}
\label{sec:amsb_plots}
\begin{figure*}[ht]
\begin{center}
\begin{tabular}{cc}
\includegraphics[width=0.422\textwidth]{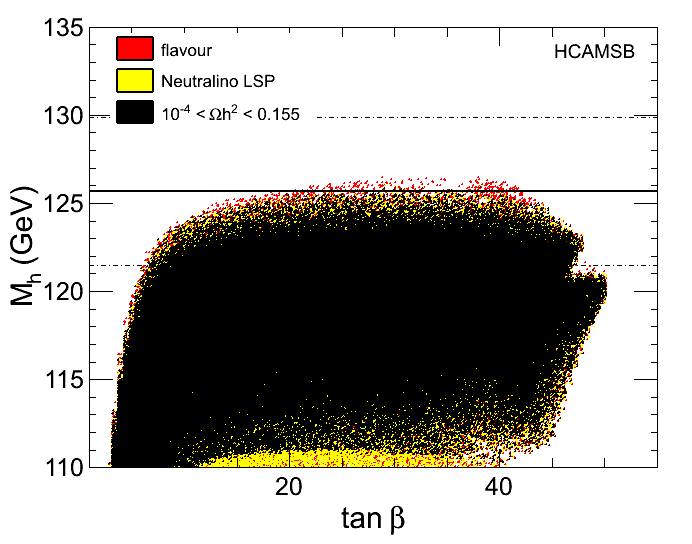}&
\includegraphics[width=0.422\textwidth]{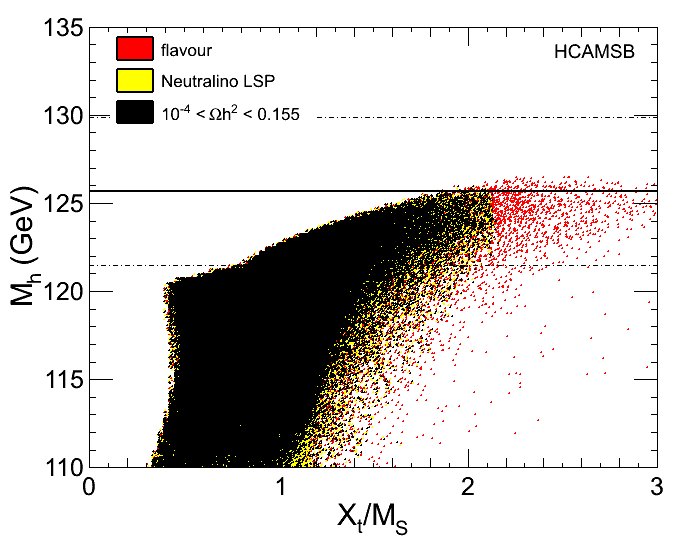}
\end{tabular}
\end{center}
\caption{\label{mh0_hcamsb}Light Higgs mass as functions of $\tan\beta$ (left panel) and $X_t/M_S$ (right panel) in HC-AMSB. The red points are all points compatible with the constraints from flavour physics described in Sec.~\ref{sec:flavour}, the yellow points have a neutralino LSP and the black points are in addition consistent with the upper bound of the relic density constraint. The horizontal solid line corresponds to the central value of the Higgs mass and the dashed lines to the 2$\sigma$ deviations. 
}
\end{figure*}
To study the mAMSB scenario, we perform flat scans by varying the parameters in the following ranges:
\begin{eqnarray}
&&m_0 \in [50,10000]\;{\rm GeV}\\
&&m_{3/2} \in [0,500]\;{\rm TeV}\nonumber\\
&&\tan\beta \in [1,55]\;,\nonumber
\end{eqnarray}
and use a sample of more than 1 million points.
We first consider the constraints obtained from the Higgs mass measurement. In Fig.~\ref{mh0_amsb}, we present the light CP-even 
Higgs mass as a function of $\tan\beta$ and $X_t/M_S$, and show the points {\em compatible} with the flavour and relic density constraints. 
First, we see that $M_h$ is limited to values below 122 GeV. 
The reason for this behaviour is related to the 
fact that in the mAMSB scenario, $X_t/M_S$ is small, corresponding to a no mixing regime which leads to a lower Higgs mass. 
Second, no mAMSB point is compatible with the tight relic density interval of Eq.~(\ref{Oh2_tight}), but there exist points compatible 
with the loose relic density interval of Eq.~(\ref{Oh2_loose}). One of the limiting factors for the light CP-even Higgs mass comes from 
the restriction $M_S < 3$ TeV that we impose to limit fine-tuning.
We have checked our results for the Higgs mass numerically using four generators: 
{\tt ISAJET} \cite{isajet}, {\tt SOFTSUSY} \cite{softsusy}, {\tt SuSpect} \cite{suspect} and {\tt SPheno} \cite{spheno}. While the results from the first three generators are fully consistent, the results of {\tt SPheno} were 
different and this may explain the different result found in \cite{Fuks:2011dg}. Our result is consistent with the one 
in \cite{Arbey:2011ab}. 

We consider now the Higgs signal strengths in Fig.~\ref{BR_tanb_amsb} as a function of $\tan\beta$. We include the 2$\sigma$ 
constraint from the Higgs mass on the plots. We first notice that most of the valid points are close to the SM values of the Higgs 
strengths. Concerning the $\mu_{\gamma\gamma}$ signal strength, ATLAS and CMS have different central values, as indicated after 
Table \ref{tab:input} and even if at 
present the average of the two values can be just used as a rough guideline, future more precise measurement are important for this class of models as values close to the SM results are favourable, while values larger than one are clearly disfavoured in 
these scenarios.
Moreover, all the Higgs strengths can be decreased, which corresponds to a suppression in the production 
cross-sections. In particular for  the Higgs to diphoton decay, the predicted strength already stands below the 2$\sigma$ experimental 
lower bound. We see however that points not compatible with the cosmology constraints can have an increased signal 
in $\gamma\gamma$  for $\tan\beta\sim20$. 
However, all these points correspond to a scenario in which the lightest supersymmetric particle (LSP) is a stau, 
and the increase is induced by light stau loops as described in~\cite{light_stau}. Nevertheless, scenarios with charged LSP are strongly 
disfavoured by the cosmology requirements for a neutral dark matter stable particle.
As a consequence, the mAMSB scenario is compatible with the Higgs mass measurements only marginally at the two-sigma level since the maximum attainable Higgs mass is below 122 GeV, and also the relic abundance constraint can only be met with the loose bounds described above.

\subsection{HC-AMSB}
\label{sec:hcamsb_plots}
\begin{figure*}[!t]
\begin{center}
\begin{tabular}{cc}
\includegraphics[width=0.422\textwidth]{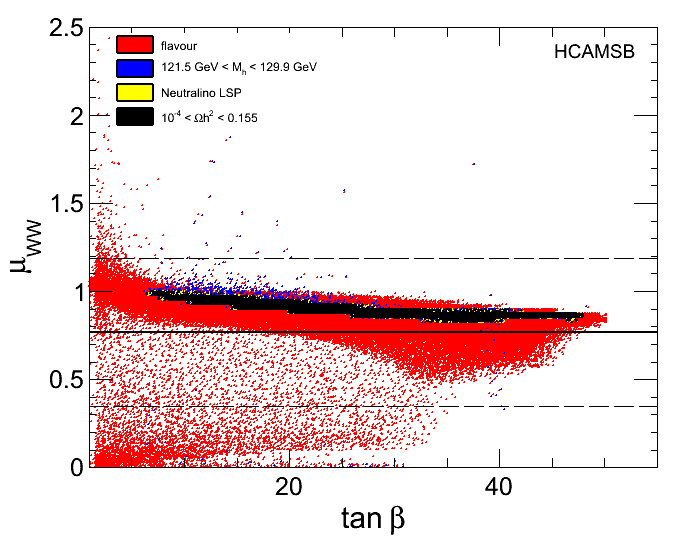}&
\includegraphics[width=0.422\textwidth]{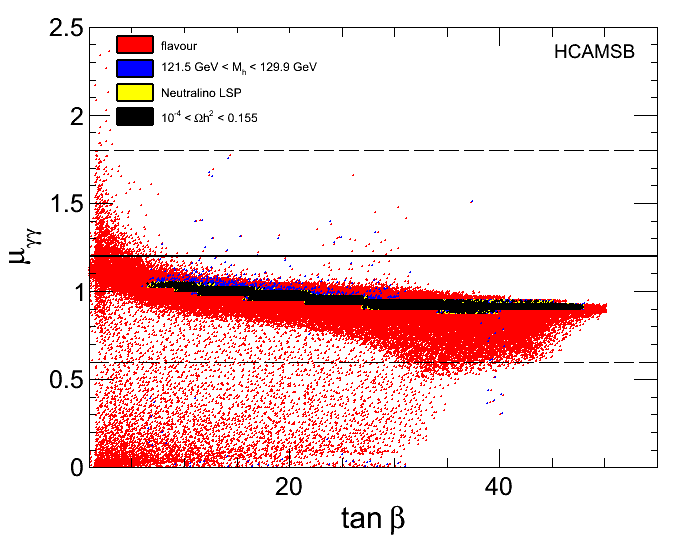}\\
\includegraphics[width=0.422\textwidth]{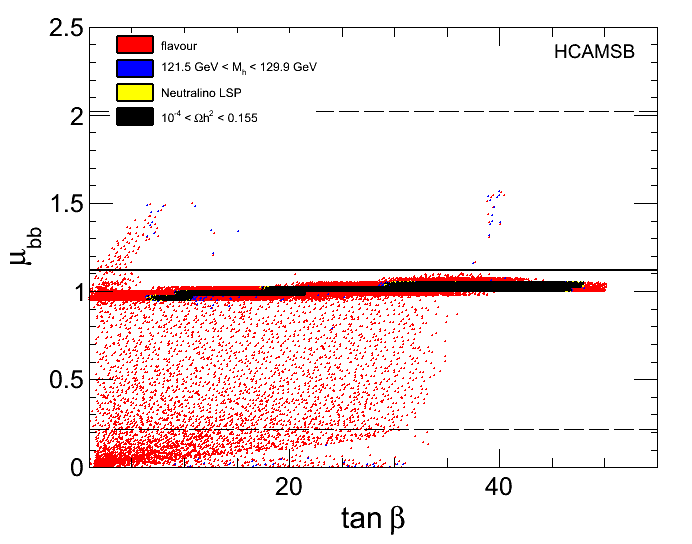}&
\includegraphics[width=0.422\textwidth]{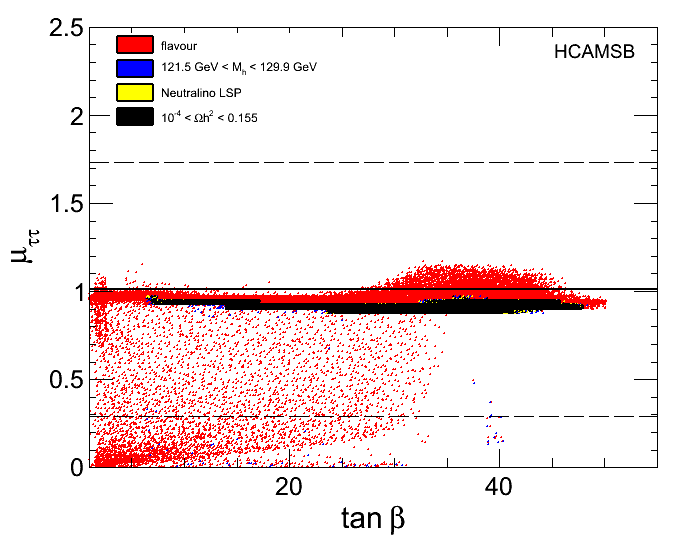}
\end{tabular}
\end{center}
\caption{\label{BR_tanb_hcamsb}$\mu_{WW}$ (upper left), $\mu_{\gamma\gamma}$ (upper right), $\mu_{b\bar{b}}$ (lower left) and $\mu_{\tau\tau}$ (lower right) as functions of $\tan\beta$ in the HC-AMSB model. The red points are favoured by the flavour physics constraints, the blue points are compatible with the Higgs mass constraint, the yellow points have a neutralino LSP and the black points in addition are compatible with the upper bound of the relic density constraint. The horizontal solid lines correspond to the experimental central values given in Table~\ref{tab:input} and the dashed lines to the $2\sigma$ intervals.}
\end{figure*}
The HC-AMSB scenario provides a modification of the $M_1$ bino mass, as discussed in Sec.~\ref{sec:model_HC-AMSB}. We 
have generated a sample of more than 1 million points through flat scans over the parameters in the following intervals:
\begin{eqnarray}
&&\alpha \in [-0.3,0.3]\\
&&m_{3/2} \in [0,500]\;{\rm TeV}\nonumber\\
&&\tan\beta \in [1,55]\;.\nonumber
\end{eqnarray}
In Fig.~\ref{mh0_hcamsb}, we plot the light Higgs mass as functions of $\tan\beta$ and $X_t/M_S$. Contrary to the mAMSB scenario, 
the Higgs mass can reach 126 GeV and therefore be fully consistent with the mass constraint. 
The sfermions are lighter in this scenario as compared to in the mAMSB scenario, as can be seen by comparing formula 
\ref{msfmamsb} and \ref{msfhcamsb} which differ by a term $m_0^2$. Numerically, as $m_0$ can typically be in the TeV range, the sfermion masses in the HC-AMSB scenario are typically lighter by the same amount with respect to the corresponding sfermions in the mAMSB model.
Moreover $X_t/M_S$ can reach larger values. On the other hand, no point in this scenario is at 
the same time consistent with the tight relic density constraint of Eq.~(\ref{Oh2_tight}), but many points fulfil both the Higgs and loose 
relic density bounds. More specifically, the allowed points have $\tan\beta \gtrsim 5$ and $X_t \gtrsim M_S$, and therefore correspond 
to the typical or maximal mixing regimes.

In Fig. \ref{BR_tanb_hcamsb}, we consider the $\mu_{WW}$, $\mu_{\gamma\gamma}$, $\mu_{b\bar{b}}$ and $\mu_{\tau\tau}$ signal 
strengths of the Higgs as a function of $\tan\beta$. First, the bulk of points compatible with the flavour constraints are consistent with the 
SM signal strengths. When imposing the Higgs mass constraint, $\tan\beta$ is restricted to large values, as already noticed in 
Fig.~\ref{mh0_hcamsb}, and most of the points with low signal strengths are removed. Finally, we impose the neutralino LSP and loose relic density constraints, and note that this requirement removes points with enhanced $\gamma\gamma$ signal strength. While this scenario is well compatible with the latest Higgs search results, it may be disfavoured in the future if the 
$\gamma\gamma$ signal strength value is confirmed to be larger than the SM value.
Thus, the HC-AMSB model can explain 
simultaneously flavour physics, loose relic density bounds and the current Higgs search results, but can be challenged by future more precise data.

\begin{figure*}[!t]
\begin{center}
\begin{tabular}{cc}
\includegraphics[width=0.422\textwidth]{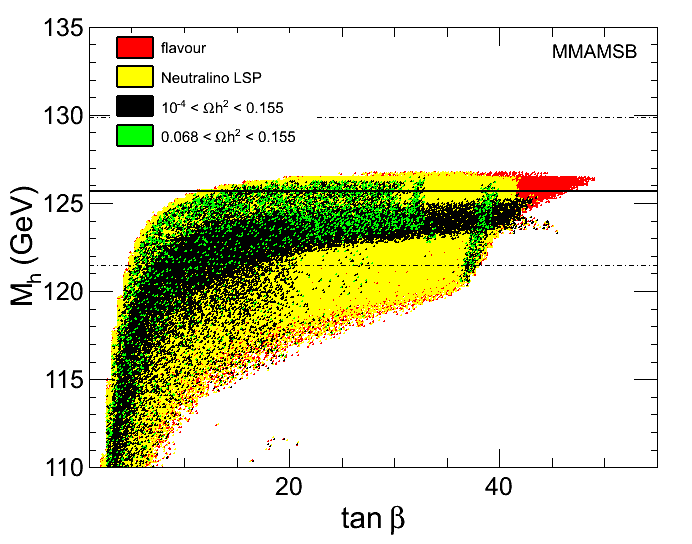}&
\includegraphics[width=0.422\textwidth]{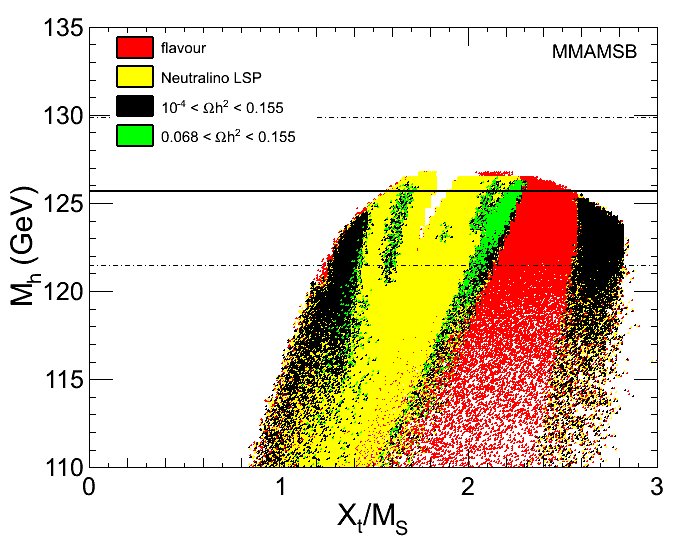}
\end{tabular}
\end{center}
\caption{\label{mh0_mmamsb}Light Higgs mass as functions of $\tan\beta$ (left panel) and $X_t/M_S$ (right panel) in MM-AMSB. The 
red points are all points compatible with the constraints from flavour physics described in Sec.~\ref{sec:flavour}. The yellow points have also a neutralino LSP. The black points are consistent with the loose relic density constraint of Eq.~(\ref{Oh2_loose}). The green points are in addition consistent with the tight relic density constraint given in Eq.~(\ref{Oh2_tight}). The horizontal solid line corresponds to the 
central value of the Higgs mass and the dashed lines to the 2$\sigma$ deviations.}
\begin{center}
\begin{tabular}{cc}
\includegraphics[width=0.422\textwidth]{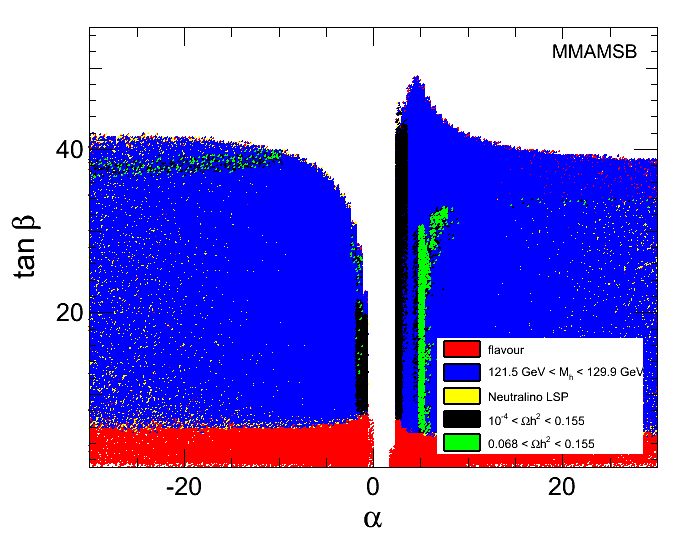}&
\includegraphics[width=0.422\textwidth]{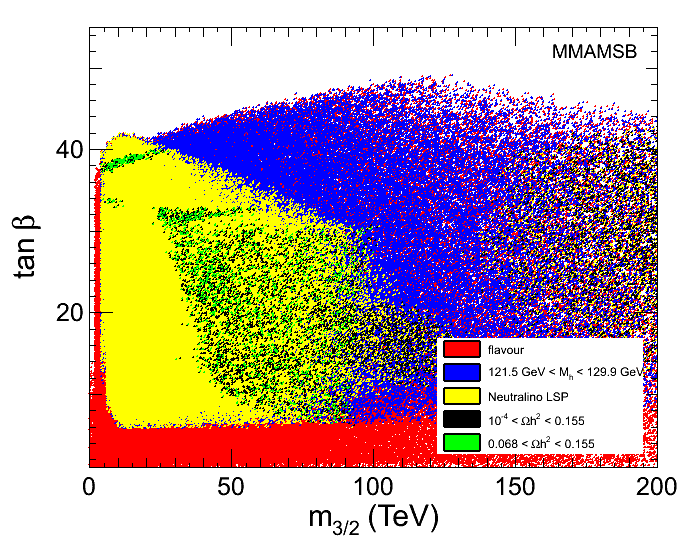}
\end{tabular}
\end{center}
\caption{\label{tanb_mmamsb}Constraints from flavour physics, Higgs mass and relic density in the $(\alpha,\tan\beta)$ (left panel) and 
$(m_{3/2},\tan\beta)$ (right panel) parameter planes in the MM-AMSB model. The red points are favoured by the flavour physics 
constraints, the blue points are compatible with the Higgs mass constraint, the yellow points have a neutralino LSP, the black points are compatible with the loose relic density constraint and the green points are in addition compatible with 
the tight relic density constraint.}
\end{figure*}
\begin{figure*}[!t]
\begin{center}
\begin{tabular}{cc}
\includegraphics[width=0.422\textwidth]{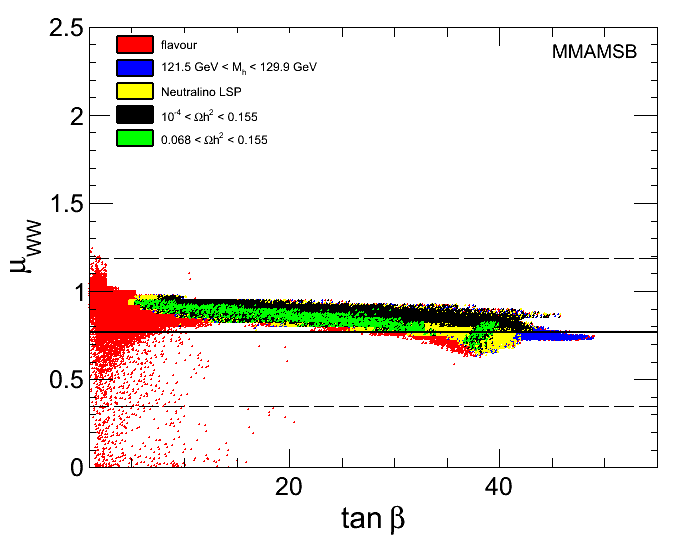}&
\includegraphics[width=0.422\textwidth]{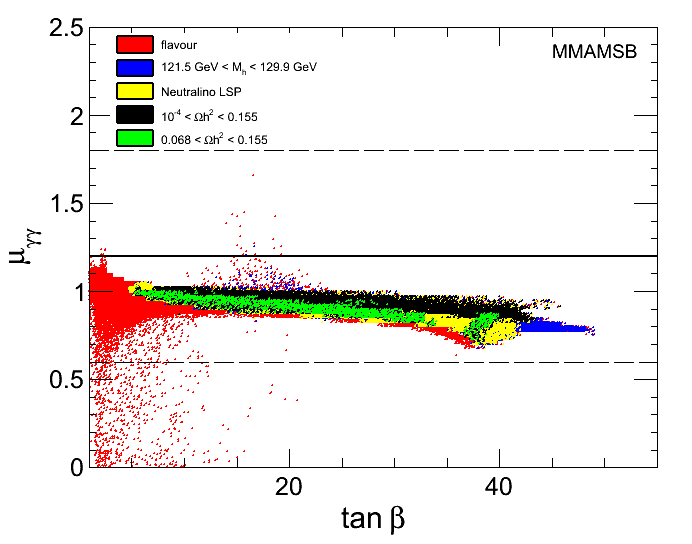}\\
\includegraphics[width=0.422\textwidth]{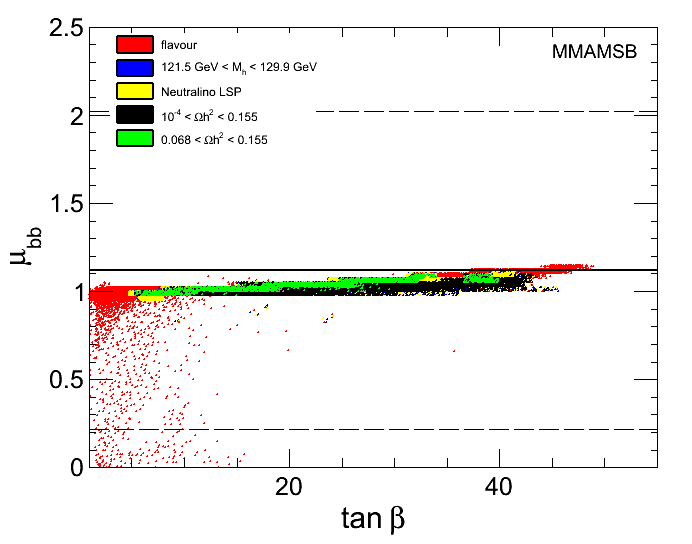}&
\includegraphics[width=0.422\textwidth]{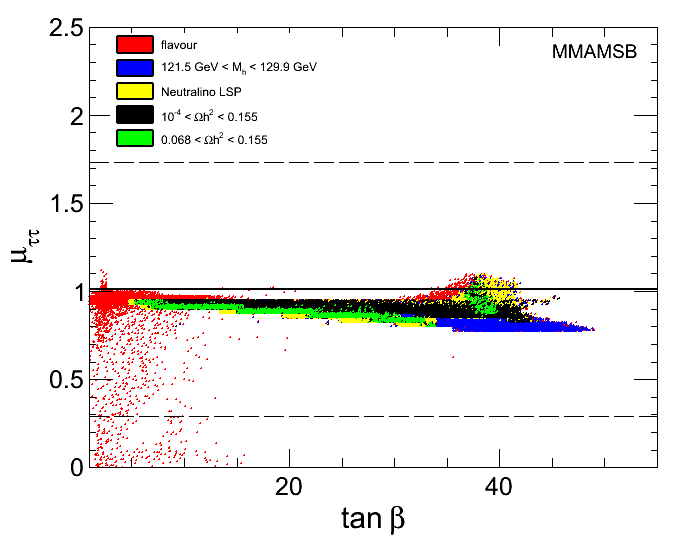}
\end{tabular}
\end{center}
\caption{\label{BR_tanb_mmamsb2}$\mu_{WW}$ (upper left), $\mu_{\gamma\gamma}$ (upper right), $\mu_{b\bar{b}}$ (lower left) and $\mu_{\tau\tau}$ (lower right) as functions of $\tan\beta$ in the MM-AMSB model. The red points are favoured by the flavour physics constraints, the blue points are compatible with the Higgs mass constraint, the yellow points have a neutralino LSP, the black points are compatible with the loose relic density constraint and the green points in addition are compatible with the tight relic density constraint. The horizontal solid lines correspond to the experimental central values given in Table~\ref{tab:input} and the dashed lines to the $2\sigma$ intervals.}
\end{figure*}

\subsection{MM-AMSB}
\label{sec:mmamsb_plots}

As we already showed in \cite{Arbey:2011gu}, the MM-AMSB has the advantage of providing solutions consistent with the tight relic 
density constraint. We confront here this model to the latest Higgs constraints. To study this scenario, we vary the parameters in the following ranges:
\begin{eqnarray}
&&\alpha \in [-30,30]\\
&&m_{3/2} \in [0,500]\;{\rm TeV}\nonumber\\ 
&&\tan\beta \in [1,55]\;,\nonumber
\end{eqnarray}
using flat scans generating more than 1 million points.

In Fig.~\ref{mh0_mmamsb}, we plot the light Higgs mass as functions of $\tan\beta$ and $X_t/M_S$. As for the HC-AMSB scenario, the 
Higgs mass can reach 126 GeV, in a region corresponding to typical and maximal mixing regimes in the stop sector. In this scenario, both the sfermion masses and trilinear couplings are modified by the modulus mediation. We note that 
imposing the lower bound of the relic density constraint makes apparent two distinct regions of compatibility: a large one with 
$\tan\beta\lesssim30$ and $X_t/M_S \gtrsim 1-2$ corresponding to a typical mixing, and a narrow strip around $\tan\beta\sim37$ and 
$X_t\gtrsim2 M_S$ corresponding to a maximal mixing.
In Fig.~\ref{tanb_mmamsb} we consider the effects of the constraints in the $(\alpha,\tan\beta)$ and $(m_{3/2},\tan\beta)$ parameter 
planes. We see clearly the difference between the two regions highlighted in Fig.~\ref{mh0_mmamsb}: the low $\tan\beta$ region has 
positive values of $\alpha$ typically around 6, while the $\tan\beta\sim37$ strip corresponds to negative $\alpha$ and small $m_{3/2}$. 
In terms of physical spectra, in both scenarios the neutralino is relatively heavy ($\gtrsim 500$ GeV). The negative $\alpha$ region 
corresponds to Higgs resonances, with a bino-like neutralino 1 mass approximately half the $H$ and $A$ Higgs masses, while the positive $\alpha$ region has stau and stop masses close to the neutralino mass, resulting in important 
co-annihilations, and the neutralino 1 is a mixed bino-wino state.

In Fig. \ref{BR_tanb_mmamsb2}, we consider the $\mu_{WW}$, $\mu_{\gamma\gamma}$, $\mu_{b\bar{b}}$ and $\mu_{\tau\tau}$ 
signal strengths of the Higgs as a function of $\tan\beta$. In comparison with the other AMSB scenarios, we find for the MM-AMSB 
model a situation similar to the one of the HC-AMSB model, where the Higgs mass constraint is satisfied, the signal strength 
for the decay of the Higgs boson to two photons is consistent with the preferred dark matter region of the parameter space within two 
sigmas.

\section{Conclusions}
\label{sec:conclusion}
Anomaly mediation and its extensions including hypercharge and moduli for supersymmetry breaking are attractive models from the 
theoretical point of view. The well-known shortcomings of these models have been largely discussed and corrected in the literature. 
However detailed phenomenological implications of the recent dark matter, Higgs, flavour and collider data were not yet considered. In 
this paper we have discussed these limits, taking into account the most important recent flavour and Higgs search results, together with 
the dark matter constraints in order to establish, which among these models are still compatible with data. 

The minimal AMSB model is consistent with the loose relic density dark matter constraints, but consistency is only marginal at the 
two-sigma level, especially due to the Higgs mass constraint. We consider therefore this minimal scenario much less attractive, once the phenomenological constraints are imposed.

Concerning the HC-AMSB model, it is consistent with the loose relic density dark matter constraints and with the Higgs mass value. Relaxing the neutralino LSP requirement and the relic density constraints allows for points with increased 
$\mu_{\gamma\gamma}$ in the region of light stau masses. This scenario with light staus has been thoroughly studied in the literature, 
however in the HC-AMSB scenario it corresponds to a region in which the stau is the LSP, making it inconsistent with cosmology. 
Contrary to the mAMSB and the HC-AMSB, the MM-AMSB model provides solutions compatible with flavour, collider data and 
the full relic density constraint. Therefore, the MM- and, to a lesser extent, the HC-AMSB model, are still attractive solutions of 
supersymmetry breaking which are consistent with present data. Future improvements in the precision of the Higgs mass 
measurements may easily rule out the minimal AMSB model if the present central value is confirmed. The MM- and HC-AMSB models 
will be still consistent in that case, but further constraints can be obtained from more precise determinations of the signal strength in the 
measured decay channels. In particular for  the $\mu_{\gamma\gamma}$ signal strength, ATLAS and CMS have currently quite 
different central values, as indicated after Table \ref{tab:input}. At present the average of the two values can be just used as a rough 
guideline. This shows the importance of present and future LHC data, in combination with flavour and dark matter constraints to 
suggest the path to be followed in the investigation of physics beyond the standard model. 

\subsection*{Acknowledgements}
We acknowledge partial support from the Labex-LIO (Lyon Institute of Origins), CF-Theorie IN2P3 and the European Union FP7 ITN INVISIBLES Network (Marie Curie Actions, PITN-GA-
2011-289442). We are also grateful to the Grenoble 125 GeV Higgs workshop where part of this work was carried out.\\
\\


\end{document}